# Restoring Execution Environments of Jupyter Notebooks

Jiawei Wang
Faculty of Information Technology
Monash University
Melbourne, Australia
jiawei.wang1@monash.edu

Li Li[α]
Faculty of Information Technology
Monash University
Melbourne, Australia
li.li@monash.edu

Andreas Zeller
CISPA Helmholtz Center for Information Security
Saarbrücken, Germany
zeller@cispa.saarland

*Abstract*—More than ninety percent of published Jupyter notebooks do not state dependencies on external packages. This makes them non-executable and thus hinders reproducibility of scientific results. We present *SnifferDog*, an approach that 1) collects the APIs of Python packages and versions, creating a database of APIs; 2) analyzes notebooks to determine candidates for required packages and versions; and 3) checks which packages are required to make the notebook executable (and ideally, reproduce its stored results). In its evaluation, we show that *SnifferDog* precisely restores execution environments for the largest majority of notebooks, making them immediately executable for end users.

## I. INTRODUCTION

Jupyter notebooks—interactive documents that combine code, text, mathematics, plots, and rich media—have become a prime medium for scientists to document, replicate, and illustrate their findings. In contrast to a regular scientific paper, a notebook allows its writers to directly interact with data and code, updating tables and diagrams on the spot. This also extends to users, who can re-execute the notebook code, say with their own data or changes to the algorithms, and see how this affects the final results. This makes Jupyter notebooks one of the most promising tools to allow for widespread replication and reuse of research results.

What sounds good in theory need not be true in practice, though, and Jupyter notebooks are no exception. Recent studies [1] have shown that the vast majority of published Jupyter notebooks can only be read by users, but not re-executed. One reason is *incompleteness,* such as the raw data not being supplied; and there is not much users can do about this. However, there are also reasons for notebooks being non-executable that can be easily avoided. One reason is that notebook code cells can be executed interactively *in any order* (and data scientists happily do so); recent approaches [2] thus focus on restoring the actual order based on internal dependencies. Another important reason, however, is that Jupyter notebooks *depend on specific environments* in which they were created, such as specific libraries in specific versions.

In principle, Python code in notebooks provides *import* statements, which state the (external) modules are to be used.

[α]Corresponding author

However, Python users install *packages,* not modules; and the names of imported modules may be different from the name of the package that provides them. Different versions of packages may provide different APIs; hence one has to determine compatible versions. Also, packages may depend on other tools or packages to be installed.

This is why Python (like other languages), in good Software Engineering tradition, has long introduced *explicit means to specify dependencies* between libraries and packages. Python package managers (e.g., pip and conda), for instance, expect Python packages to provide an explicit *list of dependencies,* stating which other packages need to be installed in which versions. Writers of Jupyter notebooks, however, are first and foremost data scientists and not software engineers [3]; hence, they neither know about principles of reusable software, nor would this be in their focus. Indeed, as we show in this paper, *around 94% of notebooks do not formally state or document dependencies;* among those who do, nearly 30% are not reliable. In consequence, users who want to execute published and complete Jupyter notebooks will very likely face errors of missing packages or incompatible versions.

In this paper, we introduce a novel approach to *automatically restore the experimental dependencies of Jupyter notebooks.* Our *SnifferDog* tool takes a Python Jupyter notebook and automatically detects which packages are required to reproduce notebook results. To this end, *SnifferDog* creates an *API bank,* a database which holds API information for each Python library (and each version). By analyzing the Python code embedded in the notebook, *SnifferDog* then determines library candidates that would be API compatible. *SnifferDog* then can automatically install the recommended dependencies and check if they allow the notebook to 1) be executed and 2) reproduce the original results stored in the notebook. When users thus apply *SnifferDog* on a notebook, they at least obtain a list of detected required libraries and their versions. If these are complete, the notebook can become executable; and in the ideal case, the notebook is shown to fully reproduce the original results. Striving for executability, reproduction, and considering library versions is also what sets *SnifferDog* apart from earlier, Python-specific approaches [4].

*SnifferDog* is efficient and effective: It finishes the analysis of 5,000 notebooks in 18,141.29 seconds (3.63 seconds per notebook). In an experiment with 315 notebooks known to be executable, *SnifferDog* was able to automatically determine dependencies for over 90% of them.

The remainder of this paper is organized as follows. After providing background about Python packages and Jupyter notebooks (Section II), we make the following contributions:

- **A study on the prevalence of dependency issues in Jupyter notebooks** (Section III). In a preliminary study, we investigated causes that make Jupyter notebooks non-executable, with and without environmental dependencies.
- **A novel approach to restore dependencies of Jupyter notebooks** (Section IV). We present the design of our approach and its implementation in the *SnifferDog* prototype.
- **An evaluation of our approach** (Section V). We evaluate the effectiveness of *SnifferDog* on a variety of notebooks, showing that it precisely restores execution environments for the largest majority of notebooks.

After discussing related work (Section VI), we close with conclusion and future work (Section VII).

## II. BACKGROUND

We start with discussing background knowledge, including Python libraries and Jupyter notebooks.

### A. Python Libraries

Python is well-known for its immense ecosystem, providing more than 200,000 third-party packages (also known as libraries) to developers. Such Python libraries need to be locally installed onto developers' implementation environment before being accessed. The Python Packaging Authority team officially maintains a standard package management tool called *pip*, which allows users to install these libraries from different sources (PyPI) [5]. In addition to package management systems, Python developers can also install a library from its source code project.

Figure 1 illustrates a typical code structure example of a Python library. A Python file named *setup.py* installs the library locally. However, this file will not install the library's environmental dependencies and, hence, requires its users to fulfill them beforehand. A *top-level package* giving the library its name (i.e., pandas) is stored in the same level of *setup.py*. Such a *package* in Python is a directory that contains a specific file named *\_\_init\_\_.py* responsible for initializing the package. Python packages provide a way to structure Python's module namespace, offering an easy means for library users to access its APIs. A *module* in Python is a specific term used in Python to specify Python's source code (i.e., files containing Python definitions and statements). Each Python file represents a Python module; for instance, *setup.py* defines a Python module named setup. Directories under the top-level package are the library's *sub-packages*. Similarly, the Python files under sub-packages are deemed as sub-modules. For example, as shown in Figure 1, the Python file *base.py* is defined as a sub-module named *base* in sub-package *pandas.io.excel*.

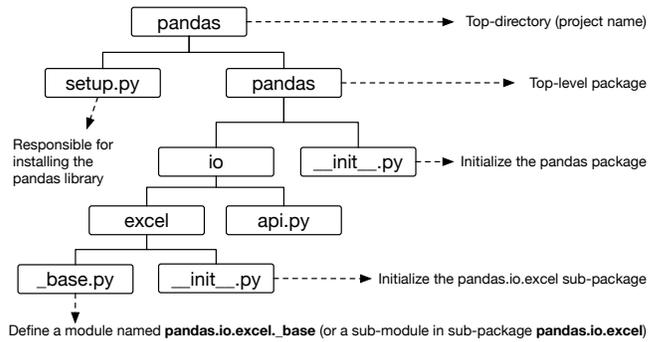

Fig. 1: A typical code structure example of Python libraries. This partial code structure is extracted from a popular Python library called Pandas.

In each Python module (or Python file), a set of methods can be declared and implemented. These methods can be accessed by other modules (or module users) and hence are referred to as *APIs*. For example, the *pandas.io.excel.base* module contains an API called *read_excel()* with the *fully qualified name* being *pandas.io.excel.\_base.read_excel()*). When Python libraries evolve, their declared API sets will likely be updated. In this work, we will leverage this information to implement *SnifferDog* so as to infer environmental dependencies for Python Jupyter notebooks.

### B. Jupyter notebooks

Jupyter notebooks are sequences of *cells*, which either contain *text* (in Markdown format) or *executable code* (and its results). In *text cells*, authors describe (using Markdown and HTML for rich formatting) the objective of the notebook and the rationale behind the code presented in the following cells. In *code cells*, authors write actual programming code, most frequently Python code. Figure 2 presents a typical example of a Jupyter notebook, containing three text cells and six Python code cells.

Each code cell can be directly executed by the underline Jupyter engine, which provides the necessary computational environment such as library dependencies. The code cells in Jupyter notebooks can be executed in any order (producing errors if its prerequisites are not satisfied). After one cell is executed, Jupyter will assign an execution order aligning with its execution order. For example, the first executed cell will be marked as "In [1]", while the fourth executed cell will be marked as "In [4]". Cells can be repeatedly executed. In such a case, the latest execution counter will overwrite the previous one.

Looking closely at Figure 2, we see that the last code cell is executed before the fourth and fifth code cells. Astute readers may have also observed that there is no information (i.e., "In [5]") indicating the fifth executed code cell. This is because when code cells are repeatedly executed, the original execution counter will be overwritten by the last execution counter—called a *skip* of execution counter. Skips make it hard to reproduce the original outputs of a notebook because the skipped execution counters are not recorded at all [2].

If the execution of a code cell generates an output (text or pictures such as diagrams), the output will also be recorded and displayed in the notebook. In Figure 2, a histogram at the end is the output of the last code cell.

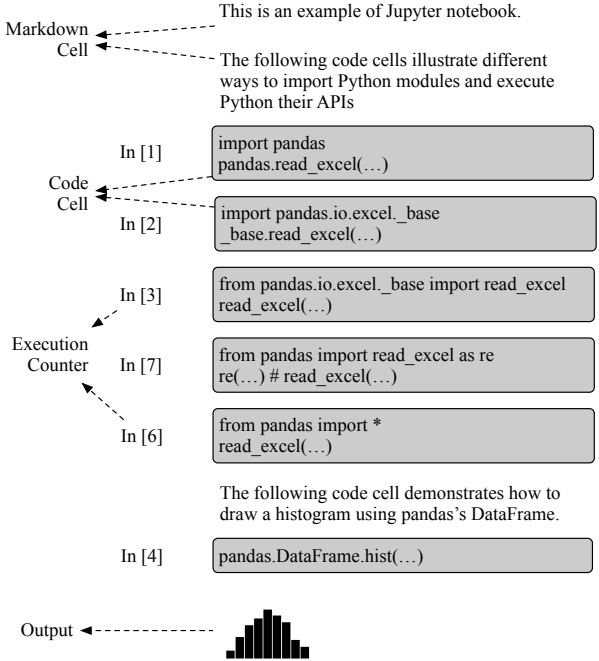

Fig. 2: An example of Jupyter notebook.

## III. PRELIMINARY STUDY AND MOTIVATION

In 2019, Pimentel et al. presented a large-scale empirical study [1] on the quality and reproducibility of Jupyter notebooks. In this study, the authors looked into 1,159,166 notebooks collected from GitHub, among which only 149,259 (roughly **12.9%**) of them were provided with module dependency information describing how the notebooks' environmental dependencies should be set up. In other words, the vast majority of existing notebooks in the community do not provide sufficient information such that notebook users could execute and replicate them. Since easy replication is one of the promises of Jupyter Notebooks, there is a need for dependable automated approaches to infer environmental dependencies for Jupyter notebooks.

How serious is this problem? We have conducted a lightweight replication study of Pimentel et al.'s work on recent Jupyter notebooks. We limit our replication study to replicating the executibility of notebooks when supplying dependencies provided by the notebook authors to identify the main causes for non-reproducibility and thus specifically address execution environments. To fulfill this purpose, we propose to answer the following research questions

- **RQ1:** To what extent do (public) Jupyter notebooks provide environment setup information?
- **RQ2:** How useful is dependency information in helping notebook users configure the execution environment?
- **RQ3:** Does the provided environment information help notebook users to execute and reproduce the notebooks? If not, what are the root causes making them non-executable?

To answer these research questions, we have collected a dataset consisting of notebooks with and without experimental setup information. Our source for notebooks is *GitHub*, one of the world's leading software repository hosting platforms. We randomly downloaded 100,000 notebooks from GitHub.

Table I summarizes our study results. Among 100,000 notebooks, less than 6% of them (or 4.74%, 1.15%, and 0.12% respectively for the three selected criteria) have been provided with environmental dependency information[1]. This rate is even lower than the rate (calculated similarly) reported by Pimentel et al. two years ago.

Note that prior work by Pimentel et al. has investigated three sources for notebook environment setup information: (1) requirements.txt, (2) Pipfile, and (3) setup.py. As discussed previously, *setup.py* is usually used to locally install a Python library from the source. It is not responsible for installing library dependencies. After manually investigating notebooks that use *setup.py,* we confirm that *setup.py* is indeed not relevant to the environment setup of Jupyter notebooks. Therefore, we exclude the third criteria *setup.py* for this study. Furthermore, when conducting the previous manual analysis, we additionally find that notebook contributors may provide environmental setup information through Anaconda (e.g., via *environments.yml*). As a result, in our study, we replace the third criteria *setup.py* with Anaconda environments.

TABLE I: Distribution of notebooks being provided with environmental dependency information w.r.t. the selected three criteria.

|  | requirements.txt | environments.yml (Anaconda) | Pipfile | Total |
| --- | --- | --- | --- | --- |
| Notebooks | 4741 | 1146 | 117 | 5826 |
| Notebooks ($\geq$3.5) | 2923 | 868 | 77 | 3740 |
| Installable | 2064 | 563 | 77 | 2646 |
| Executable | 518 | 207 | 14 | 725 |

> **RQ1:** To what extent do (public) Jupyter notebooks provide environment setup information?
>
> Among 100,000 notebooks, only less than 6% provide environmental dependency information for helping users execute their notebooks.

For the 5,826 notebooks that *have* been provided with environmental dependencies, we further check how *reliable* these are. To this end, we implemented scripts to automatically install such dependencies, using Anaconda [6] to create individual environments for each of the aforementioned Jupyter

---

[1]Some notebooks may provide two types of information for helping users setup the execution environments. For example, there are 101 notebooks contain both *requirements.txt* and Anaconda information.

```
1  ## requirements.txt ##
2  pandas     # Without specifying versions
3  scipy      == 1.17.5   # Must be version 1.17.5
4  sklearn   >=0.23.0.    # Minimum version 0.23.0
5
6  ## Conda environment.yml ##
7  name: env_name
8  dependencies:
9  - numpy=1.11.1=py35_0
10 - openssl=1.0.2h=vc14_0
11 - pandas=0.18.1=np111py35_0
12
13 ## Pipfile ##
14 [[source]]
15 url = "https://pypi.python.org/simple"
16 verify_ssl = true
17 name = "pypi"
18
19 [packages]
20 pandas = "*"
21 sqlachemy = ">= 1.3.0"
```

Fig. 3: Example of the requirement files, Anaconda environment.yml and Pipfiles.

notebooks. After that, we leveraged a tool named *nbconvert*[2] to evaluate the execution of notebooks. Since *nbconvert* in Python 3.5 or its lower versions are no longer supported by the official Jupyter team, we had to exclude 2,086 notebooks that cannot be analyzed by our scripts.

After automatically installing the remaining 3,740 notebooks, we resort to their *logs* to check whether the installations are successful or not. Based on our observation, when an installation fails, it will contain one of the following three messages: (a) "InstallationError" occur as the runtime exception, (b) "ERROR:", and (c) "cannot find a version for". Of the 3,740 notebooks, 1094 notebooks failed to have their dependency information installed, giving a failure rate of 29.25%.

> **RQ2:** How useful is dependency information in helping notebook users configure the execution environment?
>
> In our evaluation, 29.25% of the environmental dependency information provided by notebook contributors was found to be unreliable and/or insufficient.

For the notebooks that we can successfully install their environmental dependencies, we go one step deeper to check if the installed dependencies are *adequate* to support the execution of the notebooks. Recall that notebook code cells can be executed in any order and can be repeatedly executed (resulting in skipped execution counters). Hence, it is practically impossible to infer the actual execution order initially conducted by the notebook contributor. In this preliminary study, we simply execute the code cells top-down. We believe this order reflects the natural flow indicating how its contributor has attempted to implement the notebook. The execution time is set 10 minutes for every notebook.

---

[2]https://github.com/jupyter/nbconvert

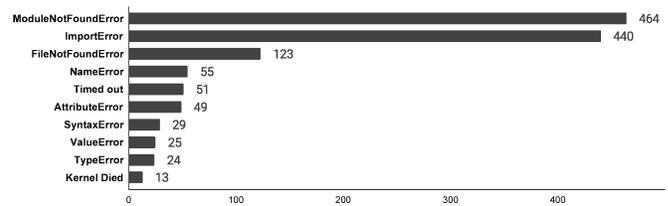

Fig. 4: Top 10 runtime exceptions from executions of notebooks whose dependency files can be successfully installed.

The last row of Table I presents the number of notebooks (i.e., Installable ones) that can be executed without errors. In other words, among 2,646 notebooks that have been provided with installable dependency information, 72.6% of them cannot be successfully executed following the straightforward top-down execution strategy.

There are various reasons causing execution of these notebooks to fail. Indeed, on the one hand, the provided dependency information may not be perfectly reliable, resulting in dependency-related errors. On the other hand, even if the environmental dependencies are correctly set up, notebooks themselves may contain implementation errors that can also lead to runtime exceptions.

Figure 4 enumerates the top-10 errors summarized from the unsuccessful notebooks. The fact that the top-ranked errors are related to environmental dependencies shows that environmental dependency seems to be the primary reason causing execution errors of the aforementioned notebooks. The top-2 ranked errors (i.e., module not found and import error) are indeed caused by the inadequate runtime environment, where the imported modules cannot be located [7]. In fact, notebooks linked to these two errors have accounted nearly half (24.15% and 22.90%, respectively) of the aforementioned unsuccessful notebooks.

```
1  # Example (1): ModuleNotFoundError
2  # from GitHub project  BenjaminBossan@mink
3  # requirements.txt
4  scikit-learn
5  ...
6
7  # Module 'sklearn.grid_search' was removed since
       scikit-learn version XXX
8  Error: No module named 'sklearn.grid_search'
9
10
11 # Example (2): ImportError
12 # From GitHub project stargaser@astrodata2016
13 # requirements.txt
14 astroquery
15 ...
16
17 # scale_image API was removed from module
       'astropy.visualization' since astroquery
       version XXX
18 Error: cannot import name 'scale_image' from
       'astropy.visualization
```

Listing 1: Two real-world examples suffering from runtime errors due to unmatched library versions.

Among various reasons causing notebooks failing to be successfully executed, we further look into some of the failures related to the top-2 types of errors and find that a significant amount of failures are due to different versions of libraries are installed. Indeed, when providing environmental dependencies (cf.Figure 3), notebook contributors are not required to specify the *exact versions* of the dependent libraries. As a result, the unmatched library versions might be installed and hence lead to runtime errors. Listing 1 demonstrates two of such examples (respectively for *ModuleNotFoundError* and *ImportError*) obtained from real-world notebooks. Due to the fast-evolving nature of software systems such as Python libraries, certain APIs might be deprecated and subsequently removed. If the wrong library versions are used, client applications, if not changed, will likely be subject to compatibility issues and hence result in runtime errors. Therefore, we argue that when specifying the environmental dependencies for Jupyter notebooks, it is essential to also *clearly specify the required versions of required libraries.*

> **RQ3:** Does the provided environment information help notebook users to execute and reproduce the notebooks?
>
> For 72.6% of notebooks, the provided dependencies are not sufficient for re-executing them without errors.

## IV. *SnifferDog*

In this section, we present our approach to automatically inferring environmental dependencies for Python Jupyter notebooks, implemented in our *SnifferDog* prototype. Figure 5 summarizes the working process of *SnifferDog*, including mainly three modules: (1) Library API Mapping, (2) Library Identification and API Standardization, and (3) API Usage Analysis. We now detail these three modules, respectively.

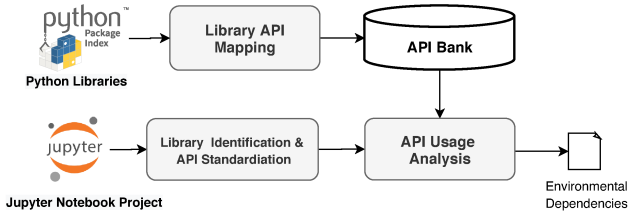

Fig. 5: The working process of *SnifferDog*.

### A. Problem Statement

Before providing the details of our approach, we formally define the problem that we plan to address in this work. At first, we need to pre-build an API bank $\mathcal{L} = \{L_1^v, L_2^v, L_3^v, ...\}$ that records a large number of popular Python libraries' API sets, in which $L_j^v$ stands for the set of APIs defined in library $L_j$ at version $v$. Then, given a Python Jupyter notebook $N$ as an input, we need to precisely parse its accessed API set $P = \{A_1, A_2, A_3, ...\}$, where $A_i$ is an API used in $P$. After that, based on the pre-built API bank $\mathcal{L}$, the goal of this approach is hence to identify a set of libraries $L$ that fulfill the following constraints: (1) $L \subset \mathcal{L}$, and (2) $\forall A_i \in P$, $\exists L_j^v \in L$ that $A_i \in L_j^v$.

### B. Library API Mapping

The first module, *library API mapping*, is not directly related to the working process of analyzing concrete Jupyter notebooks but plays an independent step in building the core infrastructure of our approach. The output of this module will be an API bank that provides an extensible (and on-growing) database recording mappings from popular Python libraries to their APIs.

Given a Python library planned to be included in the API bank, this module first builds a directory tree following the file and directory composition of the library, aiming at providing a clear way to referring library APIs (e.g. from top-level package to the leaf module). In this directory tree, Python packages (or sub-packages) are represented by non-leaf nodes and Python source code files are represented by leaf nodes. Figure 1 presents such an example, representing a partial code structure tree of the popular Python library *pandas*. This module then builds Abstract Syntax Tree (AST) trees for each leaf node (or Python file) and traverses the trees to locate public functions, including their positional and keyword parameters. The output of this step can already build a mapping from the library (in a certain version) to its defined APIs.

Unfortunately, this approach may overlook certain API usages. Applying it to *pandas* (Figure 1), the API *read_excel()* (defined in the *base.py* module) can be referenced via its full qualified name *pandas.io.excel._base.read_excel()*, or *_base.read_excel()* (or *read_excel(()*) if module *pandas.io.excel._base* (or the API itself *pandas.io.excel._base.read_excel*) is imported, as respectively shown in the second and third code cells in Figure 2. However, as demonstrated in the first, fourth, and fifth code cells Figure 2, API *read_excel()* could be invoked via another forms such as *pandas.read_excel()*, i.e., it can be directly imported from the *pandas* module despite it being defined in the *pandas.io.excel._base* module.

This ambiguity is part of the complicated Python import mechanism, which has been implemented in a transitive manner. Let $X \xrightarrow{f} Y$ be importing API $f$ from module $X$ to module $Y$ via statement *"from X import f"* in the source of $Y$. Transitivity enables $X \xrightarrow{f} Y$, if $Y \xrightarrow{f} Z$ and $X \xrightarrow{f} Z$. This feature, offered by Python runtime, has been frequently leveraged by many Python libraries to provide simplified means for users to access their APIs since it can transparently shorten the full qualified API names.

To resolve this feature, while parsing Python source code, we further conduct an *import-flow* analysis to find all the possible alternatives (or aliases) of directly defined APIs. Take the API *read_excel* again as an example, regarding the simplified source code shown in Figure 6(a), the *import-flow* analysis would lead to the following two flows.

$pandas.io.excel.\_base \xrightarrow{read\_excel} pandas.io.api$

$pandas.io.api \xrightarrow{read\_excel} pandas$

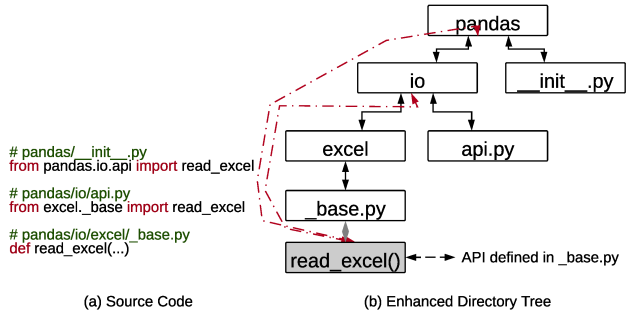

Fig. 6: An example of enhanced directory tree for Python library pandas.

By taking transitivity into consideration, we could further deduce the following flow.

$$pandas.io.excel.\_base \xrightarrow{read\_excel} pandas$$

Subsequently, at the end of this module, we further leverage these inferred and deduced import flows to enhance the directory tree initially constructed for the library (cf. Figure 6(b), before recording them into the API bank. The enhanced directory tree allows us to generate a complete list of APIs for each library integrated into the API bank.

### C. API Identification and Standardization

As shown in Figure 5, the second module *API identification* focuses on analyzing *Jupyter notebook* projects (rather than Python libraries as that being targeted by the first module) to identify and standardize library API usage. Python code in notebooks can access (1) methods available in *local modules* that are often developed by the notebooks' contributors, (2) *Python standard methods* (i.e., often known as system APIs) that are provided by the core Python modules, and (3) *library methods* (i.e., often known as library APIs) that are developed by third parties and should be imported from external resources. In this module, we are only interested in the third type of methods, namely library APIs. To distinguish library methods from local modules and system APIs, we consider all the methods that are not defined locally and are not from Pythons' system APIs as library APIs [8]–[13], [13]–[16].

Following the same approach implemented in the *library API mapping* module, we first build AST trees for the notebook's Python code and then traverse the trees to identify library APIs. After that, this step goes one step deeper to expand the identified APIs to their fully-qualified names, based on the information extracted from the *import* and *from-import* statements. For example, for the API call statement *_base.read_excel()* in the second code cell in Figure 2, the standardized API will be *pandas.io.excel._base.read_excel()*. If the identified API is an alias, i.e., *re()* in the fifth code cell in Figure 2 defined by statement *from pandas import read_excel as re*), we will further replace it with its actual name while conducting the API standardization step. The standardized version will hence be *pandas.read_excel()*. Observant readers may have observed that our approach will lead to two full-qualified names for the same API *read_excel()*. Indeed, at this stage, it is non-trivial for our approach to be aware of that by simply analyzing the notebook code. We hence consider them as two independent APIs. Nevertheless, as discussed in the previous subsection, both of these two full-qualified API names will be recorded in the API bank thanks to the *import-flow* analysis. Hence, any imprecision in the analysis will not impact the overall precision of our approach.

Moreover, Python code may involve instances of library classes that are initialized by calling constructor methods. The APIs invoked by those instances should also be appropriately identified and expanded. However, in Python, there is generally no syntax level difference between initializing constructor methods and accessing standard methods. Therefore, additional efforts are needed to distinguish them and thereby to allow the identification of classes' instances and their accessed APIs. As an example, consider the code snippet in Listing 2:

```
1 from x import y
2 m = y()
3 m.fun()
```

Listing 2: An example of qualifying object member function calls.

In Listing 2, our approach will first identify that *m.fun* is a member function call and then trace back to its construction call *m = y()*; hence method *fun()* is an API in module *x*. Subsequently, the fully qualified name of this API will be *x.y.fun*.

### D. Library Usage Analysis

The last module of *SnifferDog*, *library usage analysis* is straightforward. Based on the second module's outputs (i.e., a set of APIs), this module queries these APIs against the API bank to find possible library candidates who have provided these APIs. Normally, because each API has been provided with a full-qualified name, the API bank can often precisely locate its belonging library. The query output will hence be multiple releases (or versions) of the same library. By integrating the query results of all the identified APIs, the objective of this module is hence to find a (minimal) list of libraries and their (maximum) version ranges that cover all the identified APIs leveraged by the input notebook. *SnifferDog* will then produce its output in common formats that describe Python environmental dependencies, such *Pipfile* or *requirements.txt*.

## V. EVALUATION

To evaluate the effectiveness of *SnifferDog*, we address the following research questions:

- **RQ4:** Is *SnifferDog* effective in mapping Python libraries to their APIs?
- **RQ5:** How accurate is *SnifferDog* in inferring environment dependencies for Python Jupyter notebooks?
- **RQ6:** To what extent can *SnifferDog* assist users in reproducing Jupyter notebooks?

## A. Experimental Setting

**Dataset of Jupyter notebooks.** Recall that the goal of this work is to automatically infer environmental dependencies for Jupyter notebooks so as to help users execute and reproduce notebook outputs. To evaluate if our approach can achieve this objective, we resort to the approach introduced by Pimentel et al. [1] to collect 100,000 Jupyter notebooks from GitHub to fulfill our experiments. GitHub is the world's leading software development platform hosting millions of software repositories. The 100,000 notebooks are retrieved from GitHub projects containing files with Jupyter notebook *.ipynb* formats and declaring Python as their programming language.

**Dataset of Selected Python Libraries.** Recall that the API bank of our approach is built based on existing libraries, and it can be easily extended to include more libraries. Generally, the more libraries considered, the more comprehensive the API bank will be, and subsequently, the more precise and sound results *SnifferDog* can achieve. Since we aim at generating dependencies for as many notebooks as possible, we start by selecting the most popular 1,000 modules imported by the aforementioned 100,000 Jupyter notebooks. We then leverage PyPI, the official Python package index, to query the installation wheel files which contain the source code of library implementation these selected modules. Because several modules may belong to the same library, or some modules have not yet been indexed by PyPI, we can only locate 488 Python libraries (with 17,947 different releases) for the selected top-1000 modules. Therefore, in this work, we leverage 488 distinct Python libraries with 17,947 releases to construct the API bank.

## B. RQ4: Effectiveness of API Bank

In this research question, we are interested in evaluating the usefulness of the API bank. From the selected 488 Python libraries, the library API mapping module extracts 1,013,718 APIs to fill the API bank. Figure 7 illustrates the distribution of the number of APIs in each selected library, giving median and mean values at 321 and 2,281, respectively, after excluding outliers.

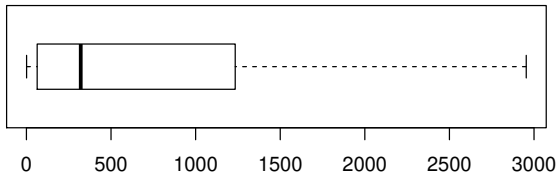

Fig. 7: The distribution of the number of APIs per library across all its version after removing outliers.

Towards evaluating the correctness of the constructed API bank, we resort to a manual process to check if these APIs are correctly recorded in the API bank. To this end, we randomly selected 166 APIs from the API bank to be manually validated. The number of selected APIs is decided by an online Sample Size Calculator [17] with a confidence level at 99% and a confidence interval at 10. For each of the selected APIs, we manually check it against its source code and find that 164 of them are correct results, giving a precision of 98.8% for our API bank construction approach.

In addition to the aforementioned manual investigation, we further resort to a *dynamic testing approach* to evaluate the correctness of the constructed API bank. Giving a mapping from a library version to its APIs, when the library (with the given version) is installed, all its APIs should be able to be imported. To this end, we implement a prototype tool to fulfill this process automatically. Specifically, we first randomly select 20 libraries, accounting for in total 3,982 APIs, from the API bank, and install them, respectively. For each of the installed libraries, we then extract all of its recorded APIs from the API bank and conduct runtime import testings to check whether these APIs can be imported at runtime. Among the 3,982 considered APIs, only 252 of them fail to be imported in our experiment, leading to a success rate of 93.6%. After analyzing the traceback information of import errors, we find that most of such failures are related to missing dependencies that are further required by the libraries under evaluation.

> **RQ4:** (Effectiveness of library mapping) Is *SnifferDog* effective in mapping Python libraries to their APIs?
>
> The API bank constructed by the Library API mapping module is precise: 98.8% of APIs are correctly extracted; 93.6% can be successfully imported.

Among the 1,013,718 APIs inferred from the 488 libraries, 686,915 of them could further introduce compatibility issues to their client applications (if incorrect library versions are installed), resulting in, for example, module not found errors and import errors. The incompatible APIs include 543,387 (53.60%) newly added APIs after the first libraries' releases, 345,234 (34.06%) removed APIs compared to the libraries' latest versions, and 58,594 (5.78%) APIs have their parameters changed over the libraries' evolution. Figure 8 further presents the distribution of newly added, removed, and updated APIs in each of the considered libraries, respectively.

Given the fact that 67.76% of the APIs (including added, removed, and updated ones without duplication) may introduce compatibility issues, there is a strong need to also infer the correct versions of the dependent libraries when inferring the environmental dependencies for Jupyter notebooks, Our API bank records the detailed evolution changes of considered libraries and is designed to infer not only the dependent libraries, but also their correct versions.

> **RQ4** (Usefulness of API bank) Is *SnifferDog* effective in mapping Python libraries to their APIs?
>
> In our evaluation, more than half of the library APIs were added, removed, or updated at some point in the libraries' life cycles. This underlines the need to check for compatible library versions, as *SnifferDog* does.

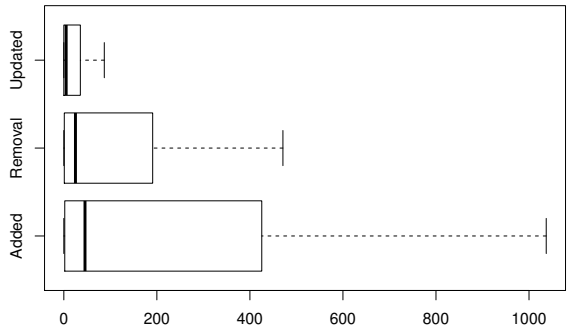

Fig. 8: The median values for the number of added and removed and updated APIs are 45, 25 and 5 respectively after excluding outliers.

*C. RQ5: Effectiveness of SnifferDog*

Let us now evaluate the effectiveness of *SnifferDog* in inferring environmental dependencies for Jupyter notebooks. We evaluate the effectiveness through one *in-the-lab* and one *in-the-field* experiment.

*1) In-the-lab experiment:* Recall that our preliminary study has identified 725 notebooks that are (1) provided with installable required dependencies, and (2) demonstrated to be executable after the provided dependencies are installed. We hence take these 725 notebooks as the ground truth to fulfill our in-the-lab experiment (because these notebooks are known executable). Unfortunately, 385 (out of the 725) notebooks have accessed libraries that are not yet considered by the current API bank (constructed based on around 488 libraries). Therefore, we have to exclude them from the ground truth. Our final ground truth is hence made up of 340 Jupyter notebooks and their required libraries.

For the 340 notebooks, we then apply *SnifferDog* to automatically generate experimental dependencies for them. After that, we follow the same approach (as discussed in Section III) to automatically install the generated libraries and execute the corresponding notebooks. Experimental results show that *SnifferDog* can successfully generate installation requirements for 315 (92.65%) notebooks, among which 284 are successfuly executed, giving a recall rate at 83.52%.

The installation failures are mainly related to library compatibility issues brought by the selected Python version (which is usually not provided by notebook contributors) and the underline Python setuptools [18]. For the 31 non-executable cases, our manual investigation reveals that the failures (8 ImportError, 7 ModuleNotFoundError and 16 other type of rumtime errors) are caused by inaccurate version constraints yielded by *SnifferDog*.

> **RQ5:** (in-the-lab) How accurate is *SnifferDog* in inferring environment dependencies for Python Jupyter notebooks?
>
> In a lab setting, *SnifferDog* is effective in automatically inferring execution environments for Jupyter notebooks, successfully generating installation requirements for 315/340 (92.6%) of notebooks. 284/315 (90.2%) of notebooks could be executed automatically.

*2) In-the-field experiment:* In this setting, we randomly select 5,000 notebooks and launch *SnifferDog* to generate execution environments for them. *SnifferDog* completes its analysis in 18,141.29 seconds, or 3.63 seconds per notebook on average.

We now check to which extent the generated environments support the execution of notebooks. To reduce human influence to a minimum, we restrict ourselves to a subset of notebooks to fulfill this purpose as it is time-consuming to evaluate a notebook, which involves installing all the dependencies and executing all of its code cells. To this end, we apply the following inclusion criteria to retain notebooks that (1) have been provided with pre-defined dependencies, which, however, cannot support their executions, and (2) are within the capacity of our API bank. This gives us 722 notebooks for the in-the-wide experiment.

Among the 722 notebooks, *SnifferDog* can successfully generate installable dependencies for 667 of them, among which 223 of them can further lead to successful executions of the corresponding notebooks.

Note that over half of the notebooks remain non-executable. Why is that so? Our manual analysis reveals the following two main reasons (apart from issues raised by notebooks' code qualities).

- **Reason 1:** The majority of notebooks fail to be executed because of the existence of so-called *optional dependencies*, which are not directly accessed by the notebooks (hence overlooked by *SnifferDog*) but are required by the notebooks' directly dependent libraries.
- **Reason 2:** A number of failed notebooks are due to the usage of *magic functions*, a special Jupyter notebook feature allowing the access of Python modules without following Python's syntax [19]. At the moment, magic functions are simply ignored by *SnifferDog*.

Moreover, for the failed notebooks, we further look into their error messages and compare them against that outputed from executions with their own dependencies (considered as the Baseline). In this experiment, only such notebooks that fail on both sides are considered. Figure 9 presents the comparison results. Clearly, the number of errors related to the execution environment for *SnifferDog* is significantly smaller than that of the Baseline. Oppositely, *SnifferDog* leads to more errors related to the code quality of the notebooks (e.g., FileNotFound, NameError, or HTTPError) compared to that of the Baseline. This experimental result shows that, while the notebooks fail to be executed in both environmental settings, the settings resulted from *SnifferDog* are more likely to be correct than its counterpart (new runtime exceptions can be further triggered when dependencies are supplied, e.g., for FileNotFoundError, the file is not provided). Hence, even if the notebook is not executable yet, the dependencies produced by *SnifferDog* can assist users in getting closer to their goal.

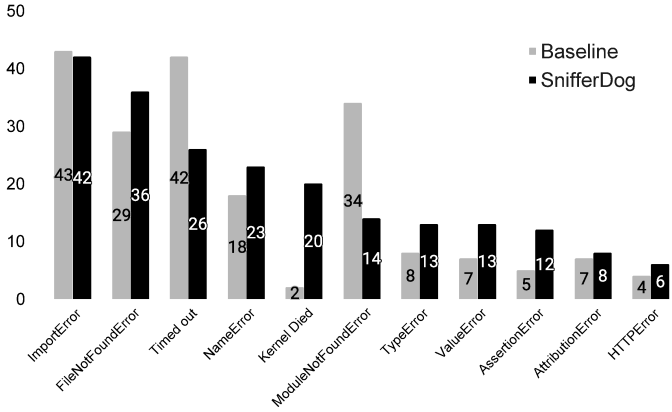

Fig. 9: Top-10 runtime error types resulted from 255 notebooks that can be not executed by environments recommended by *SnifferDog* and their original ones.

> **RQ5:** (in-the-field) How accurate is *SnifferDog* in inferring environment dependencies for Python Jupyter notebooks?
>
> Applied on random notebooks, *SnifferDog* is effective in automatically inferring execution environments. When notebooks remain non-executable, dependencies inferred by *SnifferDog* help users to get them closer to the goal.

### D. RQ6: Reproducing Jupyter notebooks

In the last research question, we evaluate if *SnifferDog* helps users not only execute, but *reproduce* Jupyter notebooks (i.e., achieving the same outputs as that recorded originally in the notebooks). To resolve the problem of cell ordering, we make use of a prototype tool called Osiris that restores the order in which cells are to be executed [2].

For our evaluation, we resort to the 507 notebooks that have been demonstrated to be executable previously. We attempt to reproduce these 507 notebooks through the following two experiments:

- **Experiment 1: Osiris (baseline).** We create execution environments for different Python versions and install all default packages (including more than 200 library packages). Then We apply Osiris in this environment to analyze the 507 notebooks.
- **Experiment 2: Osiris (*SnifferDog*).** In this experiment, we use the dependencies recommended by *SnifferDog* to set up the environment, and we launch Osiris in this environment on the same 507 notebooks.

Figure 10 presents the experimental results. Interestingly, although with over 200 popular Python libraries installed in the execution environment, around one-fourth of notebooks cannot be fully executed. Compared to the default environment, the failure rate decreases to less than 1% when the execution environment is set up based on the outputs of *SnifferDog*[3]. Regarding reproducibility, Osiris reports that, with the help of *SnifferDog*, 42 additional notebooks could be reproduced compared to the 323 notebooks reproduced with the default Osiris configuration. These experimental results show that our approach is indeed effective in helping users execute and reproduce Jupyter notebooks.

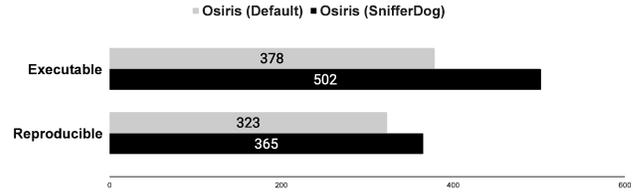

Fig. 10: Running Osiris in a default environmental setting "Osiris (Default)" vs. running Osiris in a *SnifferDog*-generated environmental setting "Osiris (*SnifferDog*)".

> **RQ6:** To what extent can *SnifferDog* assist users in reproducing Jupyter notebooks?
>
> Compared to a default environment setting, environments restored by *SnifferDog* makes significantly more notebooks executable and reproducible.

### E. Threats to Validity

The experimental findings may suffer from various threats to validity. The notebooks selected in this work may not be representative. We attempt to mitigate this threat by starting from 100,000 real-world Jupyter notebooks. We also resort to manual analysis to summarize the execution errors as well as confirm some of our experimental results. Such manual processes are known to be error-prone. We have cross-validated the results to mitigate potential errors.

## VI. Related Work

We now discuss the closely related works from three aspects: Restoring Execution Environments, Studies on Jupyter Notebooks, and Python dependency analysis.

### A. Restoring Execution Environments

The most related work to ours is *DockerizeMe* [4], inferring environment dependency configurations for Python code snippets collected from GitHub Gist. *DockerizeMe* is implemented based on a pre-acquired knowledge base of known Python packages using static analysis, which is similar to the approach of ours. However, in addition to static analysis, it further leverages dynamic analysis to achieve its purpose, being able to achieve 30% improvement in reducing *ImportError* messages. While, in principle, *DockerizeMe* could also be applied to fully analyzed Jupyter notebooks, there are important conceptual differences compared to ours. First, *DockerizeMe* only considers the latest version of packages. It hence cannot deal with programs containing deprecated, removed, renamed APIs in the latest version of packages. Indeed, as empirically reported by Horton et al. [20] by an empirical study about the executability of Python code snippets on GitHub. Their experimental results show that most gists are not executable in a default Python environment, and while a naive approach can infer dependencies for some gists, it fails to do so in the majority of cases. Second, *DockerizeMe* does not *execute* code to validate its findings, let alone compare results against published results—whereas *SnifferDog* automatically determines the configuration that makes the notebook executable, ideally even reproducing the results. The authors further proposed the tool V2 that takes the program crashes information to guide the search for correct environment dependencies [21]. However this approach relies on repeated execution of code snippets and does not handle the case when no crash happen and dependencies are incorrect. On the contrary, *SnifferDog* is a static approach that analyzes dependencies using pre-built knowledge.

---
[3]Ideally, there should be no failed cases because those notebooks have been demonstrated to be executable with the same environment. However, in practice, Osiris is implemented in Python and per se is dependent on several libraries, which may again conflict with the libraries recommended by *SnifferDog*.

### B. Studies on Jupyter Notebooks

Despite their popularity, research on Jupyter notebooks is still limited. In 2019, Pimentel et al. conducted a large scale study on the executability and reproducibility issues of over one million selected notebooks [1]. Their experimental results show that around 25% of the notebooks can be executed without any runtime errors, and among which only 4.03% of them can eventually produce the original results. Loenzen et al. [22] empirically investigate the code duplication and reuse in Jupyter notebooks and find that notebook repositories have a mean self-duplication rate of 7.6%. More recently, Wang et al [23] conduct a large scale study on the code quality and empirically find that even notable Jupyter notebooks are frequently suffered from technical debts (e.g., deprecated API uses).

Following the discovery of low-reproducibility issues among Jupyter notebooks, Wang et al. [2] went further to propose to address the root causes leading to non-reproducible notebooks by offering the community a tool called Osiris. This tool attempts to reproduce Jupyter notebooks by leveraging code instrumentation to find out and address the uncertainties when executing Jupyter notebooks. However, due to a lack of appropriate execution environments, around 80% selected notebooks failed to be fully executed. Following this research line, Fangohr et al. [24] have further proposed another tool called *nbval* (implemented as a plugin for pytest) aiming at supporting automated testing and validation of Jupyter notebooks. As argued by the authors, nbval could be leveraged to promote reproducible science such as checking that deployed software behaves as its documentation suggests.

In addition to researches from the Software Engineering community, Jupyter notebooks have been selected frequently as subjects by our fellow researchers in other domains [25]–[33]. For example, Perkel et al. [3] have studied why Jupyter notebooks are popular among data scientists. Kery et al. [26] have introduced a tool named *Verdant* to support users with efficient retrieval and sensemaking of messy version data. It allows users to compare, replay, and trace the relationships amongst different versions of artifacts of both non-code and code in the editors. Furthermore, Rule et al. [34] look into the notebooks from the aspects of human factors, and empirically observed that computational notebooks may lack the explanatory textual information.

### C. Dependency Analysis

One of the most representative ones targeting Python dependencies would be the work recently proposed by Ying et al., who attempt to resolve dependency conflicts in the Python library Ecosystem [35]. They designed and implemented a tool named *Watchman* to detect dependency conflicts among libraries indexed by the PyPI repository. They also reported 117 potential dependency issues to the developers of the corresponding projects. Despite Python has become one of the most popular programming languages nowadays, studies on Python projects focus on library API issues and their evolution patterns [36], [37], there has not been much relevant research aiming at resolving Python dependencies. Nevertheless, dependency analysis has been a hot research topic for many other programming languages [38]–[44]. The concepts of these approaches, such as resolving compatibility issues caused by the evolution of libraries [45], [46], automated replacing outdated libraries [47], or updating deprecated library APIs [48], we believe, should also be appliable to Python software applications.

## VII. CONCLUSION AND FUTURE WORK

Jupyter notebooks may be touted as a prime means to obtain reproducible and replicable research results. In practice, however, they suffer from problems that Software Engineering has solved long ago: bad code quality, insufficient documentation, and—as shown in this paper—little to nonexistent management of dependencies. It will take time until the community of Jupyter notebook authors will learn to see their notebooks not only as entities to be published, but also as living code that should be designed to be readable, reusable, and maintainable. Until then, it will be up to the Software Engineering community to reverse engineer notebooks such that they can be executed and tested.

In this paper, we have taken a major step towards this goal, namely restoring the execution environments of Jupyter notebooks. We found that dependencies are hardly ever stated explicitly, and that this problem seriously impedes re-execution of Jupyter notebooks. By analyzing imports and API usages in notebooks and matching them against Python libraries in various versions, our tool *SnifferDog* can identify library candidates that were used for notebook creation. By searching for candidate configurations that make the notebook executable again (and hence fully reproducible), *SnifferDog* provides notebook users with essential information that makes notebooks usable again. Given the popularity of notebooks, *SnifferDog* thus shows how Software Engineering can make an important contribution towards reproducible and extensible science.

There is still lots to do, though. Our future work will focus on the following topics:

**Larger API Bank.** Our API bank is only built based on 488 libraries. While these make up the most popular libraries, having a larger API Bank will further extend the capability of our approach.

**Python and C code.** A small number of "Cython" libraries combine both C and Python syntax to achieve C-like performances, letting a small set of library APIs be overlooked.

**Advanced features.** As already stated in Section V, some advanced Jupyter notebook and Python features are not yet supported by *SnifferDog*, notably magic functions and indirect dependencies.

**Beyond Python.** The *SnifferDog* principles are not limited to Python. We plan to extend *SnifferDog* to other popular Jupyter notebook languages such as R and Julia.

**Beyond notebooks.** The *SnifferDog* principles also extend beyond notebooks. *SnifferDog* could be equally applied to C source code to determine which library versions would be required for construction and execution. A version of *SnifferDog* for LaTeX that automatically determines required packages and versions may be especially welcome in scientific communities.

*SnifferDog* is available as open source (with explicit dependencies, of course). A complete *replication package* including all experimental data is available at

https://github.com/SMAT-Lab/SnifferDog.git

REFERENCES

[1] J. F. Pimentel, L. Murta, V. Braganholo, and J. Freire, "A large-scale study about quality and reproducibility of Jupyter notebooks," in *2019 IEEE/ACM 16th International Conference on Mining Software Repositories (MSR)*. IEEE, 2019, pp. 507–517.

[2] J. Wang, T.-Y. Kuo, L. Li, and A. Zeller, "Assessing and restoring reproducibility of Jupyter notebooks," in *The 35th IEEE/ACM International Conference on Automated Software Engineering (ASE 2020)*, 2020.

[3] J. M. Perkel, "Why Jupyter is data scientists' computational notebook of choice," *Nature*, vol. 563, no. 7732, pp. 145–147, 2018.

[4] E. Horton and C. Parnin, "DockerizeMe: Automatic inference of environment dependencies for Python code snippets," in *Proceedings of the 41st International Conference on Software Engineering*, ser. ICSE '19. IEEE Press, 2019, pp. 328—-338. [Online]. Available: https://doi.org/10.1109/ICSE.2019.00047

[5] "Pip user guide." [Online]. Available: https://pip.pypa.io/en/stable/user_guide/#requirements-files

[6] "Anaconda software distribution," 2020. [Online]. Available: https://docs.anaconda.com/

[7] "The Python language reference." [Online]. Available: https://docs.python.org/3/reference/import.html

[8] "Python module index." [Online]. Available: https://docs.python.org/2.7/py-modindex.html

[9] "Python module index." [Online]. Available: https://docs.python.org/3.1/modindex.html

[10] "Python module index." [Online]. Available: https://docs.python.org/3.2/py-modindex.html

[11] "Python module index." [Online]. Available: https://docs.python.org/3.3/py-modindex.html

[12] "Python module index." [Online]. Available: https://docs.python.org/3.4/py-modindex.html

[13] "Python module index." [Online]. Available: https://docs.python.org/3.5/py-modindex.html

[14] "Python module index." [Online]. Available: https://docs.python.org/3.6/py-modindex.html

[15] "Python module index." [Online]. Available: https://docs.python.org/3.7/py-modindex.html

[16] "Python module index." [Online]. Available: https://docs.python.org/3/py-modindex.html

[17] "Sample size calculator." [Online]. Available: https://www.surveysystem.com/sscalc.htm

[18] "Documentation¶." [Online]. Available: https://setuptools.readthedocs.io/en/latest/

[19] "Built-in magic commands¶." [Online]. Available: https://ipython.readthedocs.io/en/stable/interactive/magics.html

[20] E. Horton and C. Parnin, "Gistable: Evaluating the executability of python code snippets on github," in *2018 IEEE International Conference on Software Maintenance and Evolution (ICSME)*. IEEE, 2018, pp. 217–227.

[21] ——, "V2: Fast detection of configuration drift in python," in *Proceedings of the 34th IEEE/ACM International Conference on Automated Software Engineering*, ser. ASE '19. IEEE Press, 2019, p. 477–488. [Online]. Available: https://doi.org/10.1109/ASE.2019.00052

[22] A. Koenzen, N. Ernst, and M.-A. Storey, "Code duplication and reuse in jupyter notebooks," *arXiv preprint arXiv:2005.13709*, 2020.

[23] J. Wang, L. Li, and A. Zeller, "Better code, better sharing: On the need of analyzing Jupyter notebooks," in *The 42nd International Conference on Software Engineering, NIER Track (ICSE 2020)*, 2020.

[24] H. Fangohr, V. Fauske, T. Kluyver, M. Albert, O. Laslett, D. Cortés-Ortuño, M. Beg, and M. Ragan-Kelly, "Testing with jupyter notebooks: Notebook validation (nbval) plug-in for pytest," *arXiv preprint arXiv:2001.04808*, 2020.

[25] D. Koop and J. Patel, "Dataflow notebooks: encoding and tracking dependencies of cells," in *9th USENIX Workshop on the Theory and Practice of Provenance (TaPP 2017)*, 2017.

[26] M. B. Kery and B. A. Myers, "Interactions for untangling messy history in a computational notebook," in *2018 IEEE Symposium on Visual Languages and Human-Centric Computing (VL/HCC)*. IEEE, 2018, pp. 147–155.

[27] M. S. Rehman, "Towards understanding data analysis workflows using a large notebook corpus," in *Proceedings of the 2019 International Conference on Management of Data*, 2019, pp. 1841–1843.

[28] A. Rule, I. Drosos, A. Tabard, and J. D. Hollan, "Aiding collaborative reuse of computational notebooks with annotated cell folding," *Proceedings of the ACM on Human-Computer Interaction*, vol. 2, no. CSCW, pp. 1–12, 2018.

[29] S. Samuel and B. König-Ries, "ProvBook: Provenance-based semantic enrichment of interactive notebooks for reproducibility." in *International Semantic Web Conference (P&D/Industry/BlueSky)*, 2018.

[30] A. Watson, S. Bateman, and S. Ray, "PySnippet: Accelerating exploratory data analysis in Jupyter notebook through facilitated access to example code," in *EDBT/ICDT Workshops*, 2019.

[31] H. Nguyen, D. A. Case, and A. S. Rose, "NGLview–interactive molecular graphics for Jupyter notebooks," *Bioinformatics*, vol. 34, no. 7, pp. 1241–1242, 2018.

[32] H. Fangohr, M. Beg, M. Bergemann, V. Bondar, S. Brockhauser, C. Carinan, R. Costa, C. Fortmann, D. F. Marsa, G. Giovanetti *et al.*, "Data exploration and analysis with jupyter notebooks," in *17th Biennial International Conference on Accelerator and Large Experimental Physics Control Systems*, no. TALK-2020-009, 2019.

[33] M. García-Domínguez, C. Domínguez, J. Heras, E. Mata, and V. Pascual, "Jupyter notebooks for simplifying transfer learning," in *International Conference on Computer Aided Systems Theory*. Springer, 2019, pp. 215–221.

[34] A. Rule, A. Tabard, and J. D. Hollan, "Exploration and explanation in computational notebooks," in *Proceedings of the 2018 CHI Conference on Human Factors in Computing Systems*, 2018, pp. 1–12.

[35] Y. Wang, M. Wen, Y. Liu, Y. Wang, Z. Li, C. Wang, H. Yu, S.-C. Cheung, C. Xu, and Z. Zhu, "Watchman: monitoring dependency conflicts for python library ecosystem," in *Proceedings of the ACM/IEEE 42nd International Conference on Software Engineering*, 2020, pp. 125–135.

[36] J. Wang, L. Li, K. Liu, and H. Cai, "Exploring how deprecated Python library APIs are (not) handled," in *The 28th ACM Joint Meeting on European Software Engineering Conference and Symposium on the Foundations of Software Engineering (ESEC/FSE 2020)*, 2020.

[37] Z. Zhang, H. Zhu, M. Wen, Y. Tao, Y. Liu, and Y. Xiong, "How do Python framework APIs evolve? an exploratory study," in *2020 IEEE 27th International Conference on Software Analysis, Evolution and Reengineering (SANER)*. IEEE, 2020, pp. 81–92.

[38] C. Tucker, D. Shuffelton, R. Jhala, and S. Lerner, "Opium: Optimal package install/uninstall manager," in *29th International Conference on Software Engineering (ICSE'07)*. IEEE, 2007, pp. 178–188.

[39] J. Patra, P. N. Dixit, and M. Pradel, "ConflictJS: finding and understanding conflicts between JavaScript libraries," in *Proceedings of the 40th International Conference on Software Engineering*, 2018, pp. 741–751.

[40] C. Soto-Valero, A. Benelallam, N. Harrand, O. Barais, and B. Baudry, "The emergence of software diversity in Maven Central," in *2019 IEEE/ACM 16th International Conference on Mining Software Repositories (MSR)*. IEEE, 2019, pp. 333–343.

[41] Y. Wang, M. Wen, Z. Liu, R. Wu, R. Wang, B. Yang, H. Yu, Z. Zhu, and S.-C. Cheung, "Do the dependency conflicts in my project matter?" in *Proceedings of the 2018 26th ACM Joint Meeting on European Software Engineering Conference and Symposium on the Foundations of Software Engineering*, 2018, pp. 319–330.

[42] Y. Wang, M. Wen, R. Wu, Z. Liu, S. H. Tan, Z. Zhu, H. Yu, and S.-C. Cheung, "Could I have a stack trace to examine the dependency conflict issue?" in *2019 IEEE/ACM 41st International Conference on Software Engineering (ICSE)*. IEEE, 2019, pp. 572–583.

[43] L. Li, T. Riom, T. F. Bissyandé, H. Wang, J. Klein, and Y. Le Traon, "Revisiting the impact of common libraries for android-related investigations," *Journal of Systems and Software (JSS)*, 2019.

[44] X. Zhan, L. Fan, T. Liu, S. Chen, L. Li, H. Wang, Y. Xu, X. Luo, and Y. Liu, "Automated third-party library detection for android applications: Are we there yet?" in *The 35th IEEE/ACM International Conference on Automated Software Engineering (ASE 2020)*, 2020.

[45] L. Li, T. F. Bissyandé, H. Wang, and J. Klein, "Cid: Automating the detection of api-related compatibility issues in android apps," in *The ACM SIGSOFT International Symposium on Software Testing and Analysis (ISSTA 2018)*, 2018.

[46] H. Cai, Z. Zhang, L. Li, and X. Fu, "A large-scale study of application incompatibilities in android," in *The 28th ACM SIGSOFT International Symposium on Software Testing and Analysis (ISSTA 2019)*, 2019.

[47] Y. Wang, B. Chen, K. Huang, B. Shi, C. Xu, X. Peng, Y. Wu, and Y. Liu, "An empirical study of usages, updates and risks of third-party libraries in java projects," in *2020 IEEE International Conference on Software Maintenance and Evolution (ICSME)*. IEEE, 2020, pp. 35–45.

[48] L. Li, J. Gao, T. F. Bissyandé, L. Ma, X. Xia, and J. Klein, "Cda: Characterising deprecated android apis," *Empirical Software Engineering (EMSE)*, 2020.